\documentclass[%
 preprint,
superscriptaddress,
 amsmath,amssymb,
 aps,
prd,
]{revtex4-2}

\usepackage{graphicx}% Include figure files
\usepackage{dcolumn}% Align table columns on decimal point
\usepackage{bm}% bold math
\usepackage{amsmath,amssymb,amsfonts,euscript}
\usepackage{hyperref}% add hypertext capabilities
\newcommand{\deriv}[2]{\,\mbox{$\displaystyle \dfrac{{\rm d}#1}{{\rm d}#2}$}\,}

\DeclareMathOperator{\ctan}{cot}
\DeclareMathOperator{\sign}{sign}

\DeclareMathOperator{\sn}{sn}
\DeclareMathOperator{\cn}{cn}

\DeclareMathOperator{\F}{F}

\begin{document}

\preprint{APS/123-QED}

\title{Reconstruction of a star motion in the vicinity of a Schwarzschild black hole by redshift of the spectrum}% Force line breaks with \\

\author{Stanislav Komarov}
\email{StasKomarov@tut.by}
\affiliation{Theoretical Physics and Astrophysics Department, Physics Faculty, Belarusian State University, Nezavisimosti av., 4, 220030 Minsk, Belarus.}%Lines break automatically or can b
\affiliation{ICRANet-Minsk}
\author{Alexander Gorbatsievich}%
\email{Gorbatsievich@bsu.by}
\affiliation{Theoretical Physics and Astrophysics Department, Physics Faculty, Belarusian State University, Nezavisimosti av., 4, 220030 Minsk, Belarus.}

\date{\today}% It is always \today, today,

\begin{abstract}
We investigate the time evolution of the redshift of the spectrum of electromagnetic radiation that is emitted by the star that moves in the vicinity of supermassive black hole. For the case of Schwarzschild black hole we solve this problem exactly in General Theory of Relativity. Also we formulate an approach that gives possibilities to solve the inverse problem: reconstruction of motion of a star in the vicinity of black hole using the redshift data. This approach consists of two steps. The first step gives possibilities to find unique solution for the integrals of motion of the star by solution of a system of non-linear equations, rather than the direct application of standard statistical methods. For this purpose we consider properties of congruences of isotropic geodesics that connect of the worldline of the source and the worldline of the observer. The second step is the application of standard least squares method in order to obtain more accurate solution. This approach is used on the numerical model of the star in the vicinity of the Center of our Galaxy.  
%\begin{description}
%\end{description}
\end{abstract}

\maketitle

%\tableofcontents

\section{introduction}
Astrophysical observations of the stars in S-cluster that is located in the vicinity of the Galactic Center give evidences for the existence of supermassive black hole in this region \cite{Gillessen2010,Gillessen2017,Parsa2017,GC2019}. This gives possibilities to study processes that take place in the vicinity of a supermassive black hole as well as to test theories of gravity \cite{SchDetection,SchPrecession,S2in2018,S62_Star}. 

The study of electromagnetic radiation of a S-star is used for the mentioned purposes. The most appropriate characteristic of this electromagnetic radiation is the redshift of the spectrum of emitted radiation. The problem of calculation of the redshift in the framework of General Theory of Relativity was considered in many papers (see, e. g. \cite{GrouldS2,DAN2018,S2in2018,Zhang2015,Angelil,Zhang2017,GKT,GRG2018}). We will refer to this problem as direct problem. Unlike many papers, where solution of direct problem is performed using post-Newtonian approximation (see, e. g. \cite{Zhang2015,S2in2018}), our consideration uses full general relativity. Apart from this, for the solution of the direct problem it is necessary to solve boundary value problem for isotropic geodesic that connect of the source and of the observer. In many works (see, e. g. \cite{Zhang2015,GrouldS2}) this problem is solved using the tables of impact parameters, that make the accuracy of the solution no more then step of the data in the table. We solve the problem from the numerical solution of usual non-linear equation, that can be performed with much more accuracy.  

Solution of inverse problem is more interesting from the point of view of astrophysical applications. The inverse problem is the calculation of parameters of motion of the star in the vicinity of the supermassive black hole from the data of time evolution of redshift. Solution of inverse problem gives possibility to perform reconstruction of motion of the star in the vicinity of black hole. The method for the solution of the inverse problem, that is based on the direct numerical minimization of the $\chi^2$ function, is presented in \cite{GrouldS2,Zhang2015} (see also Section \ref{TheorModel}). Certain analytical results for the solution of inverse problem that are used not only redshift data, but intensity of the radiation, also is presented in \cite{Tarasenko}. 

In the present work we describe the approach that gives possibilities to solve inverse problem using data of time evolution of redshift only. Our main goal is to use analytical expressions for all quantities, describing position of the star and redshift of spectrum lines. We exclude from the obtained equations quantities, that can not be expressed through unknown parameters analytically (this is impact parameter $D$), using expressions not only for the function of redshift $z(\tau)$, but also for the derivative $\mathrm{d}z(\tau)/\mathrm{d}\tau$. Due to this solution of inverse problem is equivalent to the solution of a system of usual non-linear equations. Consideration the optical scalar $\sigma$ as a small parameter gives us possibilities to write down result equations in analytical form. As the first step of the solution we find integrals of motion of the star graphically. For the second step we find more accurate solution using least squares method.

As an example of the solution of inverse problem we consider a mathematical model of star moving in the close vicinity of a supermassive black hole. For this purpose chosen parameters of motion correspond to the star on slightly more closer orbit around black hole than known S-stars (see, e. g. \cite{GC2019,S62_Star}). This gives possibilities to test the approach for the case of strong gravitational field and gives possibilities to use the approach for the cases when the sources on very short distances to the Galactic Center, that can be found in future.        

\section{Theoretical model}
\label{TheorModel}
Consider a non-rotating, non-charging black hole. The gravitational field of such black hole can be described by Schwarzschild metric 
(see, e. g. \cite{Mi}):
\begin{equation}
\mathrm{d}s^2=\frac{\mathrm{d}r^2}{1-2M/r}+r^2\mathrm{d}\theta^2+r^2\sin^2\theta\mathrm{d}\varphi^2-(1-2M/r)\mathrm{d}t^2\,.
\end{equation} 
Here $x^i=\{t,\,r,\,\theta,\,\varphi\}$ are Schwarzschild coordinates. We use system of units such that the speed of light in vacuum $c=1$. In our model mass of the black hole $M$ (in geometrical units) is much larger than the mass of the source $M_s$. For example mass of the supermassive black hole in the Galactic Center $M\approx 4\cdot 10^{6}M_{o}$ \cite{Gillessen2017,S62_Star} and mass of the S-star is $M_s\sim M_{o}$. Because of this consider star as a test particle, moving in the external gravitational field of supermassive black hole. The components of 4-velocity of the star can be found from the geodesic equation. They are have the following form (see, e. g. \cite{Mi}):
\begin{eqnarray}\label{ueq}
&&u^1=\frac{\mathrm{d}t}{\mathrm{d}\tau}=\frac{E}{(1-2M/r)}\,;\nonumber\\
&&u^2=\frac{\mathrm{d}r}{\mathrm{d}\tau}=e_s\sqrt{E^2-(1-2M/r)(1+L^2/r^2)}\,;\nonumber\\
&&u^3=\frac{\mathrm{d}\theta}{\mathrm{d}\tau}=0\,;\nonumber\\
&&u^4=\frac{\mathrm{d}\varphi}{\mathrm{d}\tau}=\frac{L}{r^2}\,,
\end{eqnarray}  
where the orientation of the spatial part of coordinate system chosen in such way that the trajectory of the star lie in the plane $\theta=\pi/2$. $L$ is the angular momentum of the star per unit mass and $E$ is the energy of the star per unit mass. $\tau$ is the proper time of the star. Factor $e_s$ takes into account either receding or approaching part of the trajectory under consideration: 
$$e_s=\sign\left(\deriv{r}{\tau}\right).$$
From the system of equations (\ref{ueq}) trajectory of motion of the star can be found in analytical form (see, e. g. \cite{Ch}):
\begin{eqnarray}
&&\frac{1}{r}=\frac{1}{r_s(\varphi,\,\delta,\,p_1,\,p_2)}=\nonumber\\
&&\frac{1}{p_2}+\frac{p_2-p_1}{p_1p_2}\sn^2\left[(\varphi-\delta)\frac{1}{2}\sqrt{1-4M/p_2-2M/p_1},\,k_s\right]\,,\notag\\
\label{rsfun}
\end{eqnarray} 
where 
\begin{equation*}
k_s=\sqrt{\frac{p_2-p_1}{p_1p_2/2-Mp_2-2Mp_1}}\,,
\end{equation*}
$\sn[\varphi,\,k]$ --- Jacobi sinus of the first kind (see \cite{Korn} for definition). $\delta$ --- longitude of the pericenter. $p_1$ and $p_2$ are pericenter and apocenter distances respectively. They are uniquely related to $E$ and $L$ as follows:
\begin{eqnarray}
&&L=\frac{p_1p_2}{\sqrt{-p_1p_2+\dfrac{1}{2}p_1p_2(p_1+p_2)-p_1^2-p_2^2}}\,;\nonumber\\
&&E=\nonumber\\
&&\sqrt{\frac{M^2-p_1p_2+\dfrac{1}{2}(p_1+p_2)\left(4M+\dfrac{p_1p_2}{M}-p_1-p_2\right)}{M^2+\dfrac{1}{2}(p_1+p_2)\left(\dfrac{p_1p_2}{M}-p_1p_2\right)}}\,.\nonumber\\
\label{EL}
\end{eqnarray} 

Proper time of the star $\tau$ can be expressed as function of its angular coordinate $\varphi$ by well-known analytical formula (see, e. g. \cite{Ch}): 
\begin{equation}\label{taufun}
\tau=\tau_s(\varphi,\,E,\,L)\,.
\end{equation}
Due to cumbersomeness of this formula, we will not write down this expression exactly.

In modern observational astronomy, stars in the vicinity of Galactic Center are investigated using the wavelength of electromagnetic radiation that has the range of $1 \mu m - 10 m$ (including pulsars, see, e. g. \cite{Gillessen2010}). This is small compare to the scale of motion of the star. Due to this we use the geometrical optics approximation (see, e. g. \cite{Stephani}). In this approximation electromagnetic radiation propagate along a null geodesic with tangent vector $k_i$, that satisfy the following relations:  $k_jk^j=0$ and $k_{i;j}k^j=0$. Therefore, choosing coordinate frame $\tilde{K}: \{t,\,r,\,\tilde{\theta},\,\tilde{\varphi}\}$ such that the observer resides on the axis $\tilde{\theta}=0,$ $\tilde{\varphi}=0$, obtain that trajectory of light ray lie in the plane $\tilde{\varphi}=const$ and (see, e. g. \cite{Mi}):   	
\begin{eqnarray}\label{keq}
&&k^1=\frac{\mathrm{d}t}{\mathrm{d}\lambda}=\frac{1}{(1-2M/r)}\,;\nonumber\\
&&k^2=\frac{\mathrm{d}r}{\mathrm{d}\lambda}=e_r\sqrt{1-(1-2M/r)D^2/r^2}\,;\nonumber\\
&&k^3=\frac{\mathrm{d}\tilde{\theta}}{\mathrm{d}\lambda}=-\frac{D}{r^2}\,;\nonumber\\
&&k^4=\frac{\mathrm{d}\tilde{\varphi}}{\mathrm{d}\lambda}=0\,,
\end{eqnarray} 
where $\lambda$ is an affine parameter along the ray, $D$ is the impact parameter. Factor $e_s$ takes into account either receding or approaching part of the trajectory of light under consideration: 
$$e_r=\sign\left(\deriv{r}{\tau}\right).$$
The sign in the expression of derivation of $\tilde{\theta}$ (see equation (\ref{keq})) chosen in such way, that $D>0$ (we consider only trajectories of zero order (see, e. g. \cite{Tarasenko,BisnovatyiRays})). 

From (\ref{keq}) obtain the following analytical expression for the trajectory of the ray:
\begin{eqnarray}
&&\frac{1}{r}=\frac{1}{r_r(\tilde{\theta},\,D)}=\nonumber\\
&&\frac{1}{P}-\frac{Qk^2}{2PM}\cn^2\left[\frac{\tilde{\theta}}{2}\sqrt{\frac{Q}{P}}+
\F\left[\arccos{\left(\sqrt{\frac{2M}{Qk^2}}\right)},k\right],k\right]\,,\notag\\
\label{rSol}\\
\nonumber
\end{eqnarray} 
where
\begin{eqnarray}
&&Q=\sqrt{P^2+4PM-12M}\,;\\
&&k=\sqrt{\frac{Q-P+6M}{2Q}}\,,
\end{eqnarray}
$cn\left[\varphi,\,k\right]$ and $F\left[\varphi,\,k\right]$ are the Jacobi cosine and the elliptic integral of first kind respectively (see \cite{Korn} for definition). For the real values $P$ has physical meaning of closest approach distance. For the complex values $P$ has no direct physical meaning (see, e. g. \cite{Ch}), but in all cases $P$ can be expressed through the impact parameter $D$ as follows:
\begin{equation*}
P=-\frac{2}{\sqrt{3}}D\sin{\left[\frac{1}{3}\arcsin{\left(\frac{3\sqrt{3}}{D}\right)}-\frac{\pi}{3}\right]}\,.
\end{equation*}

Angular coordinates in both coordinate systems are connected by the relation:
\begin{equation}\label{angles}
\tilde{\theta}=\arccos[\cos(\varphi)\sin(i)]\,,
\end{equation}
where angle $i$ is the inclination of the orbit of the star.

Redshift of the electromagnetic spectrum can be calculated from the formula (see, e. g. \cite{Zhang2015})
\begin{equation}\label{redshiftForm}
z=\frac{\delta\lambda}{\lambda}=\frac{(u_i)_s (k^i)_s}{(u_j)_o(k^j)_o}\,.
\end{equation}
Here $\lambda$ is the wavelength of emitted light, $\delta\lambda$ is the difference between wavelength of the received light and the emitted light. $(k_i)_s$ and $(k_i)_o$ denote the wave vector in the points of radiation and receiving of radiation by the observer respectively. $(u_i)_o$ and $(u_i)_s$ denote 4-velocity vector of the observer and of the star respectively.

\begin{figure}[h]
	\includegraphics[width=7 cm]{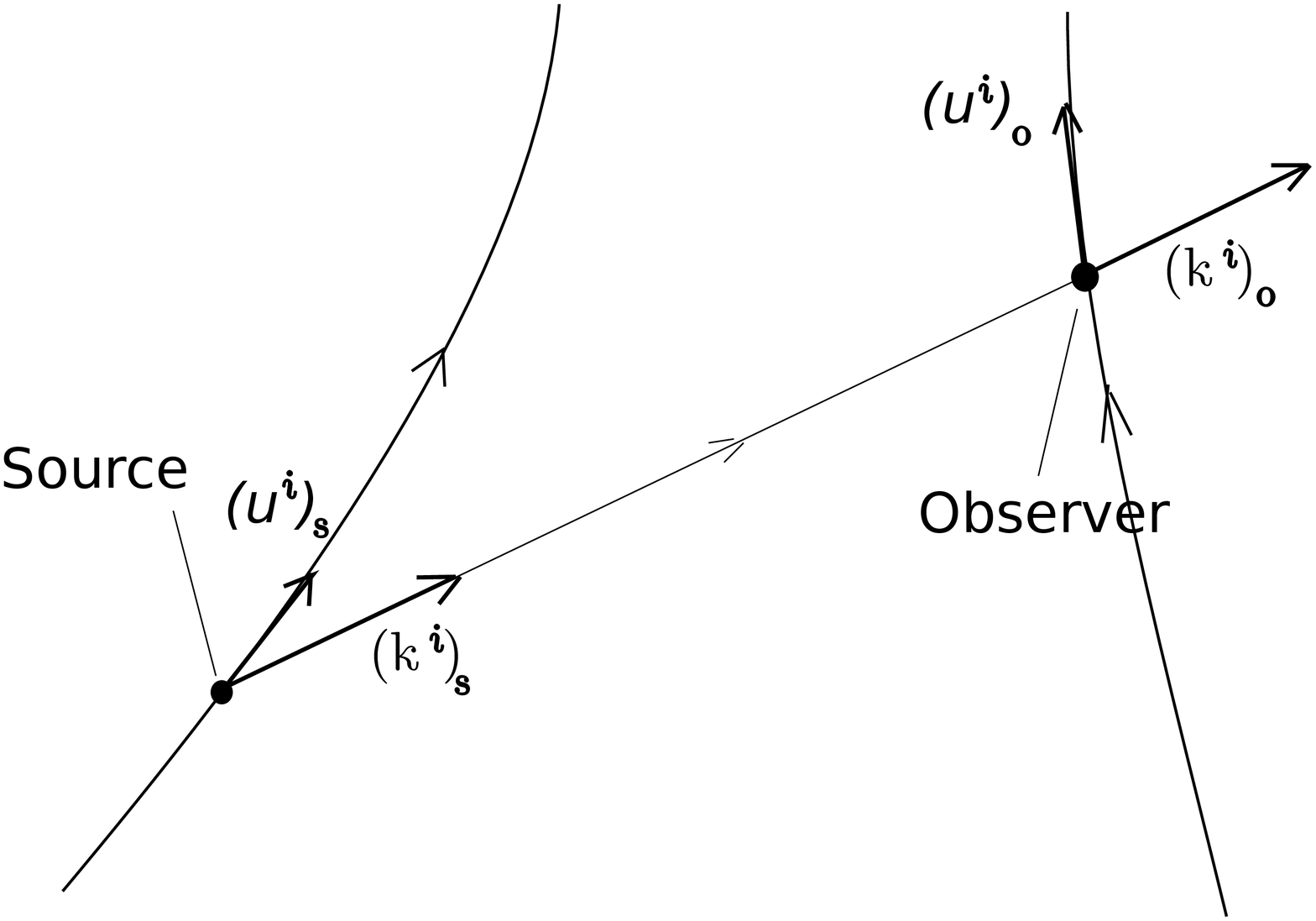}% Here is how to import EPS art
	\caption{\label{generalredshift} $(k^i)_o$ and $(k^i)_s$ are tangent vectors to null geodesic that intersect both: the worldline of the source and the worldline of the observer.}
\end{figure}

In common observational scenario observer resides far away from the source of gravitational field. Therefore metric in the vicinity of observer with good accuracy can be considered as Minkowskian metric. Motion of this observer (for example, motion of the Earth in the case of ground based observer) can be taken into account using the standard formulas (see, e. g. \cite{Damour-DeruelleII,Teukolsky}). Because of this consider only the case of stationary observer, resided on spatial infinity. This means that $(u^i)_o=\{1,\,0,\,0,\,0,\}$. The 4-velocity of the star can be found solving the equations (\ref{ueq}) numerically. But in order to calculate components of the wave vector of the ray it is necessary to solve boundary value problem for the system of differential equations (\ref{keq}) (see FIG. \ref{generalredshift}), such that the solution describes isotropic geodesics intersecting both: world line of the source and world line of the observer. In general case this problem has infinite number of solutions. In this paper we will consider only light rays of 0 order (see, e. g. \cite{Tarasenko, BisnovatyiRays}). Therefore, solution is unique. Observationally, redshift of light from 0 order trajectories can be detected as redshift of the most bright image of the star (see, e. g. \cite{BisnovatyiRays}). In the chosen abbreviations solution of the mentioned boundary value problem reduce to the solution of the following non-linear ordinary equation for the impact parameter $D$:
\begin{equation}\label{BVPeq}
r_s(\varphi,\,p_1,\,p_2)=r_r(\tilde{\theta},\,D)\,.
\end{equation}  
Solution for 0 order trajectories corresponds to the solution of (\ref{BVPeq}) with maximal value of $D$. 

It is follows from the definition of redshift (\ref{redshiftForm}), which can be rewritten in the form 
\begin{equation}\label{zform}
z=\delta t_{o}/\delta{\tau}-1\,,
\end{equation}
 where $\delta t_{o}$ and $\delta{\tau}$ are time intervals of the observer and the source between corresponding events (that connected by isotropic geodesic).  Time of observation $t_{o}$ can be found from the following relation
\begin{equation}\label{ttau}
t_{o}=\int_{0}^{\tau}(z(\tau')+1)\mathrm{d}\tau'\,
\end{equation}
where redshift  $z(\tau')$ is considered as a function of the proper time $\tau'$ of the source.  

\begin{figure}[h]
	\includegraphics[width=9 cm]{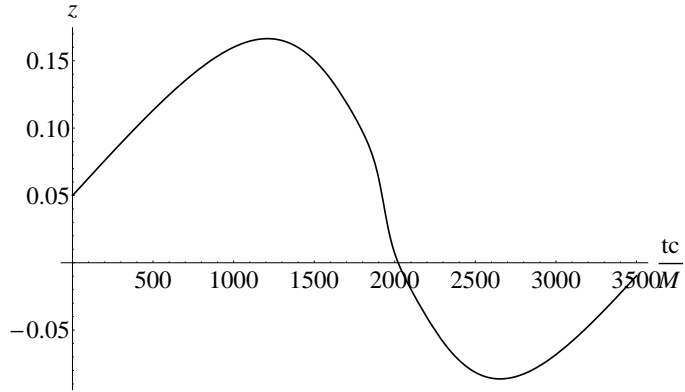}% Here is how to import EPS art
	\caption{\label{SchwarzschildRedShift2} Redshift of electromagnetic spectrum of a star in the external Schwarzschild gravitational field as function of time of observation $t$. Pericenter distance of the star orbit $p1=60 M$, apocenter distance of the star orbit $p_2=90 M$, inclination of the orbit $i=1.4\, rad$.}
\end{figure}

Take into account stationarity of the observer, relation between angles (\ref{angles}), and substitute expression (\ref{ueq}), (\ref{keq}) into (\ref{redshiftForm}). Obtain redshift in the following form:
\begin{eqnarray}\label{zSchwarzschild}
&&z=-1+\frac{E}{1-2M/r}+\frac{DL}{r^2}\beta-e_s e_r\times\nonumber\\
&&\frac{\sqrt{\left(E^2-\left(1-\dfrac{2M}{r}\right)\left(1+\dfrac{L^2}{r^2}\right)\right)\left(1-\left(1-\dfrac{2M}{r}\right)\dfrac{D^2}{r^2}\right)}}{1-2M/r}\,.\nonumber\\
&&
\end{eqnarray}
Here the abbreviation $\beta=\sin(i)\sin(\varphi)/\sin(\tilde{\theta})$ is introduced. The presented equations give possibilities to solve the direct problem: calculation of redshift of electromagnetic spectrum of a star, moving in external gravitational field of supermassive black hole as function of time of observation. We illustrate the presented method by the numerical model with realistic parameters for the star, moving in the vicinity of the Galactic Center. The results of calculation are presented in FIG. \ref{SchwarzschildRedShift2}. 

However, for the astrophysical purposes it is more interesting to solve the inverse problem: determination of the parameters of motion of a star in the external gravitational field of supermassive black hole from the data of redshift of electromagnetic spectrum of the star. In literature the inverse problem solved by minimization of $\chi^2$ function (see, e. g. \cite{GrouldS2,Zhang2015}):
\begin{equation}
\chi^2=\sum\limits_{j=1}^N\left[\frac{(z_j-z_{\text{obs}, j})^2}{\sigma^2_Z}\right]\,,
\end{equation}
where $z_j$ and $z_{\text{obs}, j}$ are the theoretical and the observation values of redshift respectively for the time of observation $t_j$ ($j\in[1,\,N]$). $\sigma^2_Z$ is the dispersion of the observation redshift data. Due to the function $z_{\text{obs}, j}(t)$ has no analytical expression (because of $D$ is the solution of non-linear equation (\ref{BVPeq})), the minimization of $\chi^2$ can be performed only numerically.

In the present work we present another approach, that is based on the derivation of the system of equation for the parameters of motion in analytical form. In order to obtain equations for inverse problem in algebraic form it is necessary to find expressions not only for $z$ but also for $\mathrm{d}z/\mathrm{d}\tau$. This calculation is described in the following section. 
    
\section{Derivative of redshift function}
\subsection{Newman-Penrose null tetrad and optical scalars}
Introduce the following Newman-Penrose null tetrad (see, e. g. \cite{PenroseRindler,N-PCoefficients}) along the world line of the ray of electromagnetic radiation of the star:
\begin{equation}\label{NPtetrad}
k^i,\, n^i,\, m^i\, \bar{m}^i\,.
\end{equation}
Here $k^i$ is the wave vector of the ray. The symbol $\,\bar{}\,$ denotes the complex conjugation. All vectors of (\ref{NPtetrad}) are null. All scalar products between vectors  (\ref{NPtetrad}) are equal to 0 apart from
\begin{equation}
k_in^i=-1;\,m_i\bar{m}^i=1\,.
\end{equation} 
Consider congruence of isotropic geodesics that have tangent vectors $k^i$ and intersect the world line of the observer in the time $t_{o}$. Also consider the Newman-Penrose tetrad (\ref{NPtetrad}) in all points of this congruence. Then the components of the vectors of the tetrad in the coordinate basis of $\tilde{K}$ have the form (we chose the affine parameter $\lambda$ such that $k_0$=-1):
\begin{eqnarray}
&&k_j=\left\{-1,\,e_r\frac{\sqrt{1-(1-2M/r)D^2/r^2}}{1-2M/r},\,-D,\,0\right\}\,;\nonumber\\
&&n_j=\left\{-\frac{1}{2}(1-2M/r),\,-\frac{e_r}{2}\sqrt{1-(1-2M/r)D^2/r^2},\right.\nonumber\\
&&\left.\frac{D}{2}(1-2M/r),\,0\right\}\,;\nonumber\\
&&m_j=\nonumber\\
&&\frac{1}{\sqrt{2}}\left\{i\frac{D}{r},\,i e_r r\sqrt{1-(1-2M/r)D^2/r^2},\,r\sin{\tilde{\theta},\,0}\right\}\,;\nonumber\\
&&\hat{m}_j=\nonumber\\
&&\frac{1}{\sqrt{2}}\left\{0,\,-i\frac{D}{r},\,-i e_r r\sqrt{1-(1-2M/r)D^2/r^2},\,r\sin{\tilde{\theta}}\right\}\,.\nonumber\\
\label{NPcomponents}
\end{eqnarray}
For the considered congruence one can obtain the following equations (see, e. g. \cite{N-PCoefficients,AnalyticalLensing,NovikovFrolov}, $\epsilon=k_{[i;j]}=0$ \cite{GRG2018})
\begin{eqnarray}
&&k_{i;j}m^i\bar{m}^j=-\rho\,,k_{i;j}m^im^j=-\sigma\,,\nonumber\\
&&k_{i:j}k^j=k_{i;j}k^i=0\,.
\end{eqnarray}
Here $\rho$ and $\sigma$ are optical scalars. They are obey the following system of equations (see, e. g. \cite{PenroseRindler,N-PCoefficients,AnalyticalLensing,NovikovFrolov}):
\begin{eqnarray}
&&\frac{\mathrm{d}\rho}{\mathrm{d}\lambda}=\rho^2+\sigma^2\,,\nonumber\\
&&\frac{\mathrm{d}\sigma}{\mathrm{d}\lambda}=2\sigma\rho+\Psi_0\,.
\end{eqnarray}
Here $\Psi_0=R_{ijsl}k^{i}m^{j}k^{s}m^{l}=3D^2/r^5$ \cite{AnalyticalLensing}. Chose the affine parameter of the ray $\lambda$ in such way that $\lambda=0$ in the point of radiation. $\lambda_0$ is the value of the affine parameter in the space point of the observer location. In the limit of Minkowsky space-time ($M\rightarrow 0$) obtain $\rho=-1/(\lambda-\lambda_0)$, $\sigma=0$.  Introduce the following abbreviations $\nu=\lambda_0-\lambda$, $X=-1/\rho$. Then obtain  
\begin{eqnarray}
&&\frac{\mathrm{d}X}{\mathrm{d}\nu}=-1-X^2\sigma^2\,,\nonumber\\
&&\frac{\mathrm{d}\sigma}{\mathrm{d}\nu}=\frac{2\sigma}{X}-\frac{3D^2}{r^5}\,.\label{OpticalSystem}
\end{eqnarray}	
Due to the observer is far from black hole, the initial conditions for the system of equations (\ref{OpticalSystem}) must be chosen as: 
\begin{equation}
X(0)=0,\,\sigma(0)=0\,.\label{OpticalInitial}
\end{equation}
The system of equation (\ref{OpticalSystem}) with initial conditions (\ref{OpticalInitial}) can be solved numerically. Only for the sum $\rho+\sigma$ an analytical expression can be obtained (see, e. g. \cite{AnalyticalLensing}):
\begin{eqnarray}
&&\rho+\sigma=-\frac{\dfrac{\mathrm{d}}{\mathrm{d}\lambda}\left(r\sin{\tilde{\theta}}\right)}{r\sin{\tilde{\theta}}}=\nonumber\\
&&-\frac{e_r}{r}\sqrt{1-\left(1-\frac{2M}{r}\right)\frac{D^2}{r^2}}+\frac{D}{r^2}\ctan{\tilde{\theta}}\label{analyticOptics}\,.
\end{eqnarray} 
Write down components of vector of 4-velocity of the star in the basis of null tetrad (\ref{NPcomponents}):
\begin{equation}\label{unull}
u^j=\frac{1}{\sqrt{2}}\left(\bar{A}m^j+A\bar{m}^j\right)+Bk^j+Cn^j\,.
\end{equation}
Obtain $-k_ju^j=(1+z)=C$. Denote the components of the killing vector $\frac{\partial}{\partial t}$ as $\xi^j$. Then
\begin{equation}
\xi^j=\frac{1}{2}\left(1-\frac{2M}{r}\right)k^j+n^j\,.
\end{equation}
Also obtain
\begin{equation}\label{Etetrad}
E=-u^i\xi_i=B+\frac{1}{2}\left(1-\frac{2M}{r}\right)(1+z)\,.
\end{equation}
From the relation for the norm of $u_i$ obtain
\begin{equation}\label{normtetrad}
u_ju^j=|A|^2-2(1+z)B=-c^2\,.
\end{equation}
From (\ref{Etetrad}) and (\ref{normtetrad}) follows
\begin{eqnarray}\label{coefftetrad}
&&B=\frac{c^2+|A|^2}{2(1+z)}\,,\nonumber\\
&&|A|^2=-c^2+2E(1+z)-(1-2M/r)(1+z)^2\,.\nonumber\\
\end{eqnarray}

\begin{figure}[h]
	\includegraphics[width=\columnwidth]{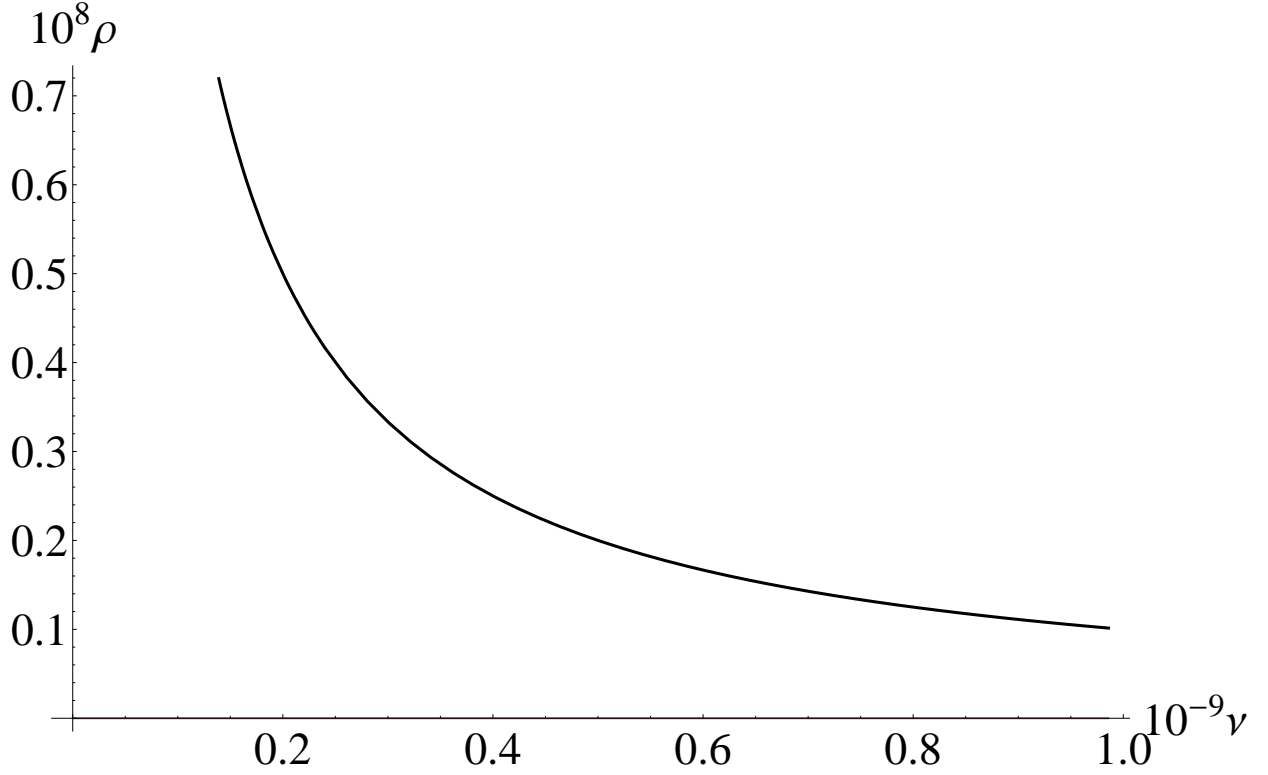}% Here is how to import EPS art
	\caption{\label{rho} $\rho$ as function of parameter $\nu$. Impact parameter $D=50M$.}
\end{figure}

\begin{figure}[h]
	\includegraphics[width=\columnwidth]{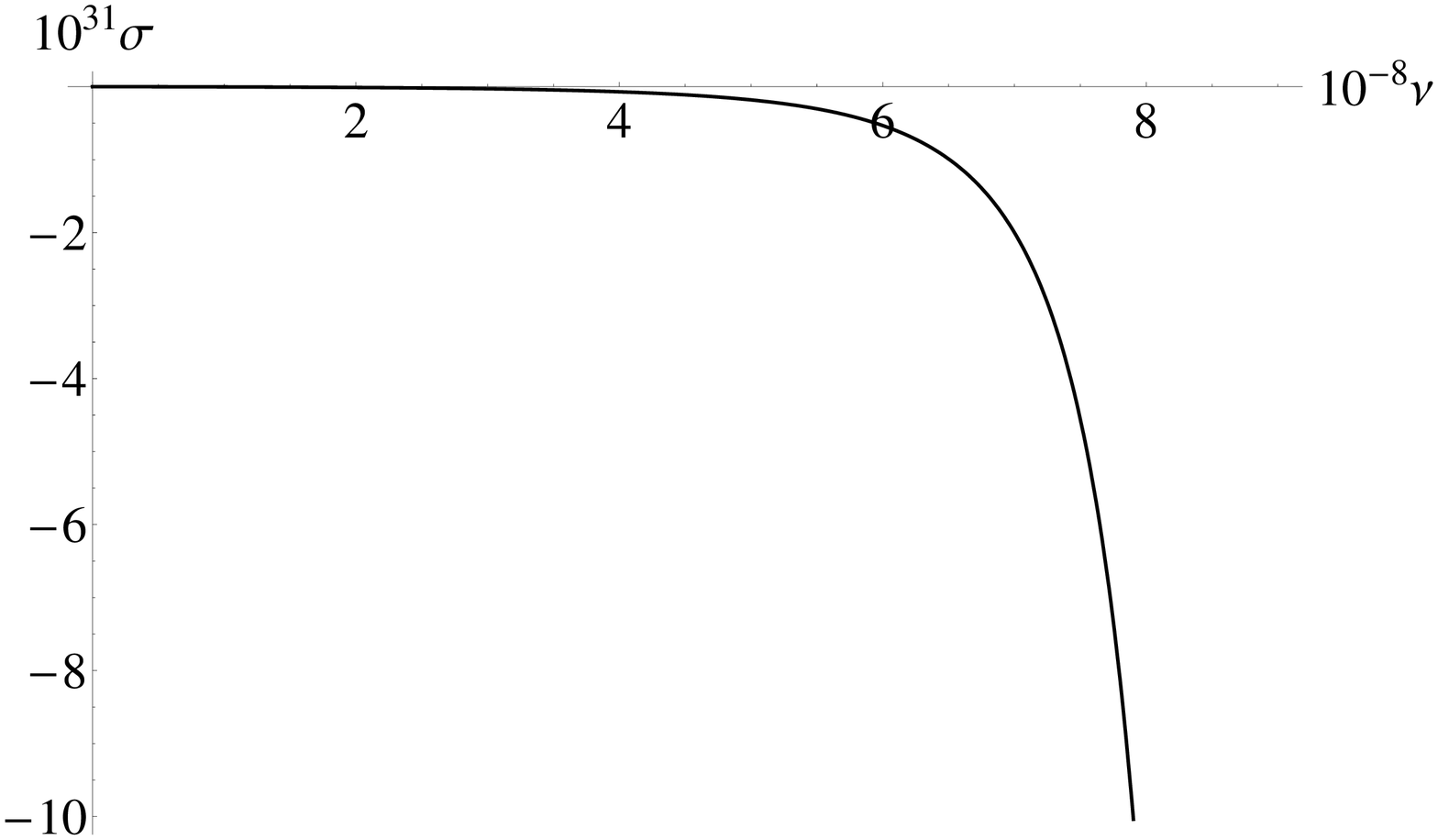}% Here is how to import EPS art
	\caption{\label{sigma1} $\sigma$ as function of parameter $\nu$. Impact parameter $D=50M$.}
\end{figure}

\begin{figure}[h]
	\includegraphics[width=\columnwidth]{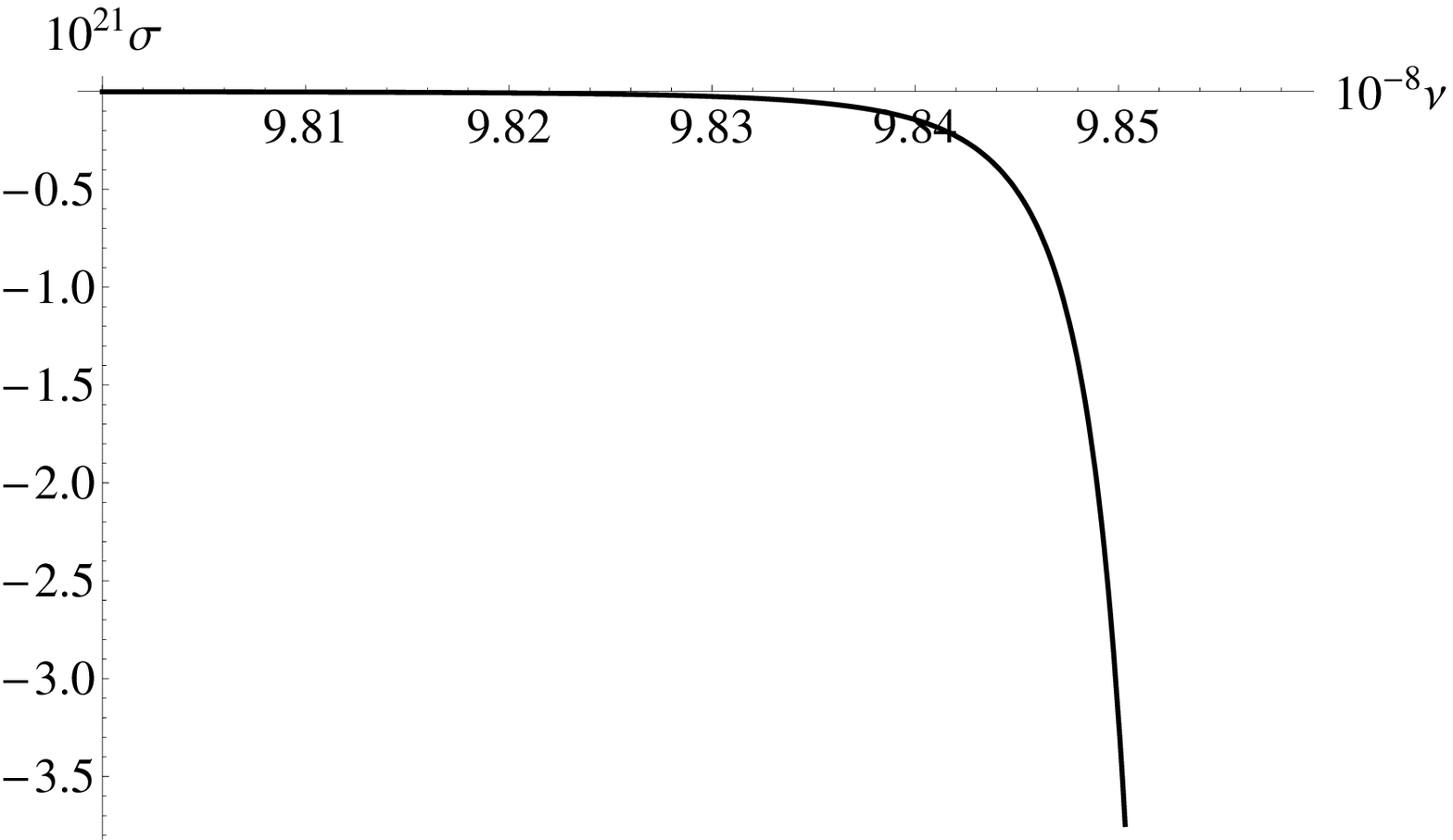}% Here is how to import EPS art
	\caption{\label{sigma2} $\sigma$ as function of parameter $\nu$. Impact parameter $D=50M$.}
\end{figure}

Now, express the time derivative of redshift, using the relation $k_{[j;l]}=0$ (see, e. g. \cite{GRG2018}):
\begin{eqnarray}
&&\frac{\mathrm{d}z}{\mathrm{d}\tau}=-k_{j;l}u^ju^l=-|A|^2 k_{j;l}m^j\bar{m}^l-A^2 k_{j;l}m^jm^l-\nonumber\\
&&\bar{A}^2 k_{j;l}\bar{m}^j\bar{m}^l-2\bar{A}(1+z)k_{j;l}n^jm^l-2 A (1+z)k_{j;l}n^j\bar{m}^l=\nonumber\\
&&|A|^2(\rho+\sigma\cos{2 P_A})-2\sqrt{2}(1+z)\frac{D}{r^3}|A|\sin{P_A}+\nonumber\\
&&\frac{e_r}{r^2}(1+z)^2\sqrt{1-\left(1-\frac{2M}{r}\right)\frac{D^2}{r^2}}\,.\label{derivz}
\end{eqnarray}
Here we use the abbreviation $A=|A|e^{i P_A}$, where $|A|$ and $P_A$ are real. It is follows from numerical calculations, that in the cases, when spatial point of radiation resided far enough from the black hole ($r\agt 20 M$), holds the following inequalities: $\rho\agt\sigma$, $\sigma<<1$ (see FIG (\ref{rho})-(\ref{sigma2}) for illustration). 
Due to this we will neglect the value of $\sigma$ in calculations. 

Then, for the time derivative of redshift obtain
\begin{eqnarray}\label{zderivativeRes}
&&\frac{\mathrm{d}z}{\mathrm{d}\tau}=\left[2E(1+z)-\left(1-\frac{2M}{r}\right)(1+z)^2-c^2\right]\times\nonumber\\
&&\left[-\frac{e_r}{r}\sqrt{1-\left(1-\frac{2M}{r}\right)\frac{D^2}{r^2}}+\frac{D}{r^2}\ctan{\hat{\theta}}\right]-\nonumber\\
&&2\frac{D}{r^3}\sin{P_A}\sqrt{2E(1+z)-\left(1-\frac{2M}{r}\right)(1+z)^2-c^2}+\nonumber\\
&&\frac{e_r}{r^2}(1+z)^2\sqrt{1-\left(1-\frac{2M}{r}\right)\frac{D^2}{r^2}}\,.
\end{eqnarray} 

In order to obtain an analytical formula, it is convenient to exclude the impact parameter $D$ from the equations (\ref{zSchwarzschild}) and (\ref{zderivativeRes}). Obtain:
%\begin{widetext}
\begin{eqnarray}
&&\frac{D}{r}=\frac{\frac{L}{r}\beta\left((1+z)\left(1-\frac{2M}{r}\right)-E\right)\pm\sqrt{\left(E^2-\left(1-\frac{2M}{r}\right)\left(1+\frac{L^2}{r^2}\right)\right)\left(|A|^2+(\beta^2-1)\frac{L^2}{r^2}\right)}}{E^2-1+\frac{2}{r}+\frac{L^2}{r^2}(\beta^2-1)(1-\frac{2}{r})}=\EuScript{F}_1(r,\,z,\,E,\,L,\,i,\,\tilde{\theta})\,;\label{1deriv}\\
&&\text{ and }\frac{D}{r}=\EuScript{F}_2(r,\,z,\,\frac{\mathrm{d}z}{\mathrm{d}\tau},\,E,\,L,\,i,\,\tilde{\theta})=\nonumber\\
&&\frac{r\frac{\mathrm{d}z}{\mathrm{d}\tau}\left(|A|^2\ctan{\tilde{\theta}}-\frac{2(1+z)}{r}|A| S\right)\pm\left(\frac{(1+z)^2}{r}-|A|^2\right)
	\sqrt{\left(1-\frac{2M}{r}\right)\left[\left(\frac{(1+z)^2}{r}-|A|^2\right)^2-r^2\left(\frac{\mathrm{d}z}{\mathrm{d}\tau}\right)^2\right]+\left(|A|^2\ctan{\tilde{\theta}}-\frac{2(1+z)}{r}|A|S\right)^2}}{\left(1-\frac{2M}{r}\right)\left(\frac{(1+z)^2}{r}-|A|^2\right)^2+\left(|A|^2\ctan{\hat{\theta}}-\frac{2(1+z)}{r}|A| S\right)^2}\,.\nonumber\\
&&\label{2deriv}
\end{eqnarray} 
%\end{widetext}
Here $S=\sin{P_A}$. An algorithm for making a chose of the sign $\pm$ in formulas (\ref{1deriv}) and (\ref{2deriv}) is given in Appendix A.   
In order to find exact expression for $\sin{P_A}$ it is possible to use the low of angular momentum conservation:
\begin{equation}\label{Ldef}
u_i\Psi^i=L=const\,,
\end{equation}
where $\Psi^i$ is Killing vector field, associated with the symmetry of the Schwarzschild metric relative to spatial rotation around arbitrary axis (chose it to be orthogonal to the orbit plane). Components of $\Psi^i$ in the coordinate basis of $\hat{K}$ have the form (see, e. g. \cite{Stephani})
\begin{equation}\label{psi}
\Psi^j=\left\{0,\,0,\,(\cos{i}+\sin{i}\ctan{\tilde{\theta}}\cos{\tilde{\varphi}}),\,\sin{i}\sin{\tilde{\varphi}}\right\}\,.
\end{equation}
Equation of the orbit plane has the following form
\begin{equation}\label{orbitplane}
-\sin{i}\sin{\tilde{\theta}}\cos{\tilde{\varphi}}+\cos{i}\cos{\tilde{\theta}}=0\,.
\end{equation}
From equations (\ref{Ldef}), (\ref{psi}), (\ref{orbitplane}), (\ref{unull}) and (\ref{NPcomponents}) follows
\begin{eqnarray}\label{sinequation}
&&e_r\frac{l}{r}\frac{1-\beta^2}{|A|\sqrt{1-\left(1-\frac{2M}{r}\right)\frac{D^2}{r^2}}}+e_P\beta\sqrt{1-\frac{l^2}{r^2}\frac{(1-\beta^2)}{|A|^2}}=\nonumber\\
&&\beta\sin{P_A}+e_r\frac{\sqrt{1-\beta^2}}{\sqrt{1-\left(1-\frac{2M}{r}\right)\frac{D^2}{r^2}}}\cos{P_A}\,,
\end{eqnarray}
where $e_P$ defined as
\begin{eqnarray}
&&e_P=\sign\left[e_r\beta l\sqrt{1-\left(1-\frac{2M}{r}\right)\frac{D^2}{r^2}}\right.+\nonumber\\
&&\left.e_s D\sqrt{E^2-\left(1-\frac{2M}{r}\right)\left(1+\frac{L^2}{r^2}\right)}\label{e_P}\right]\,.
\end{eqnarray}
The exact solution of (\ref{sinequation}) has the form
\begin{equation}\label{sinP_A}
\sin{P_A}=e_P\sqrt{1-\frac{L^2}{r^2}\frac{(1-\beta^2)}{|A|^2}}\,.
\end{equation}
\section{Inverse problem}
\subsection{Surface of integrals of motion}
\begin{figure}[h]
	\includegraphics[width=\columnwidth]{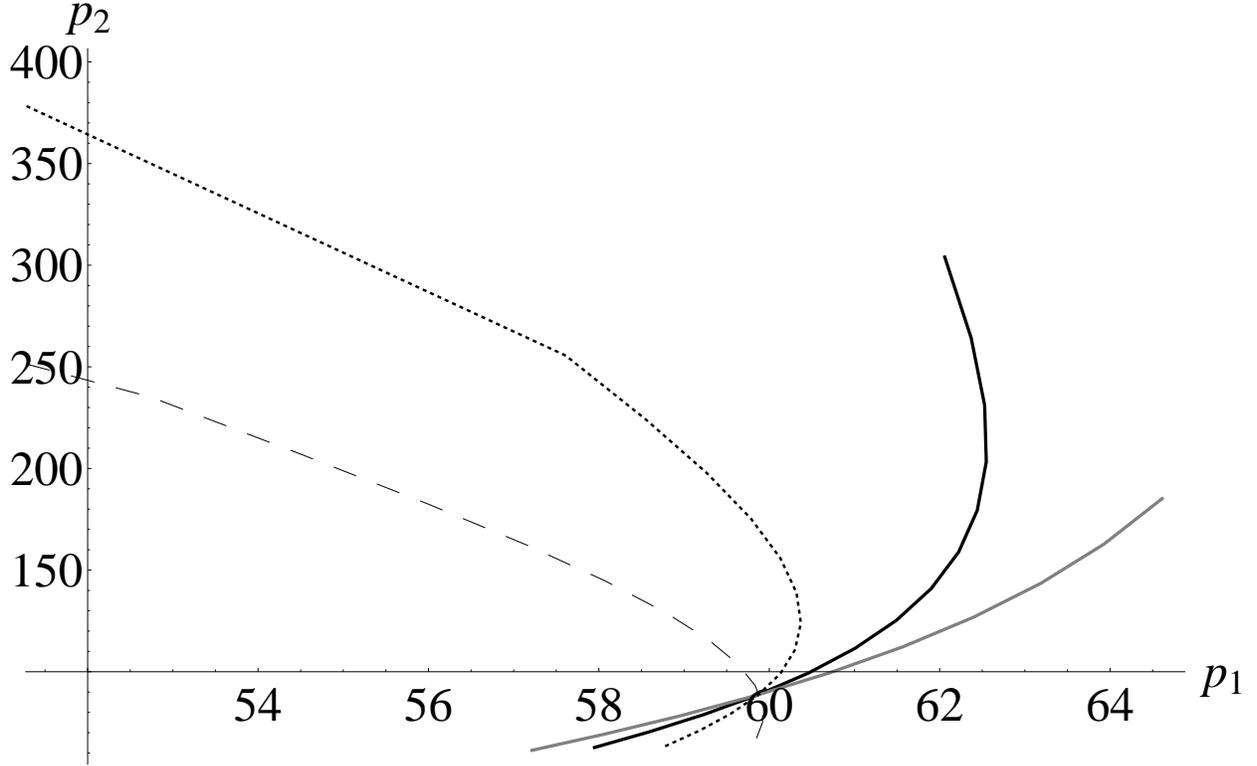}% Here is how to import EPS art
	\caption{\label{d1pi2}2-sections of solution of  (\ref{FirstEq}), (\ref{SecondEq}) for $i=\pi/2$ and $\delta=1$ by plane {$p_1$, $p_2$} for different points of data (see also Fig. \ref{SchwarzschildRedShift2}}
	\end{figure}

\begin{figure}[h]
    \includegraphics[width=\columnwidth]{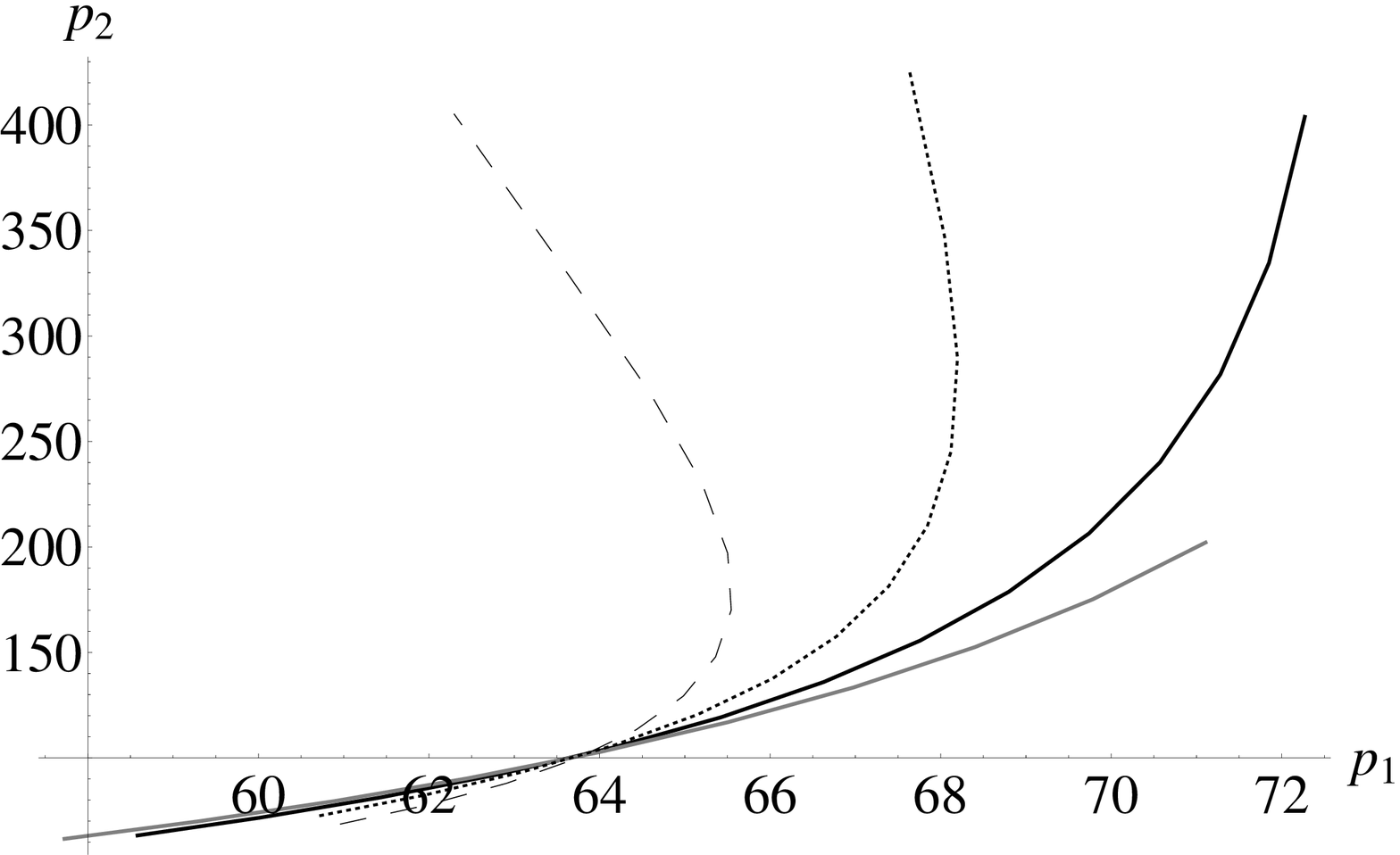}% Here is how to import EPS art
    \caption{\label{d14pi2}2-sections of solution of  (\ref{FirstEq}), (\ref{SecondEq}) for $i=\pi/2$ and $\delta=1,4$ by plane {$p_1$, $p_2$} for different points of data (see also Fig. \ref{SchwarzschildRedShift2}}
\end{figure}

\begin{figure}[h]
	\includegraphics[width=\columnwidth]{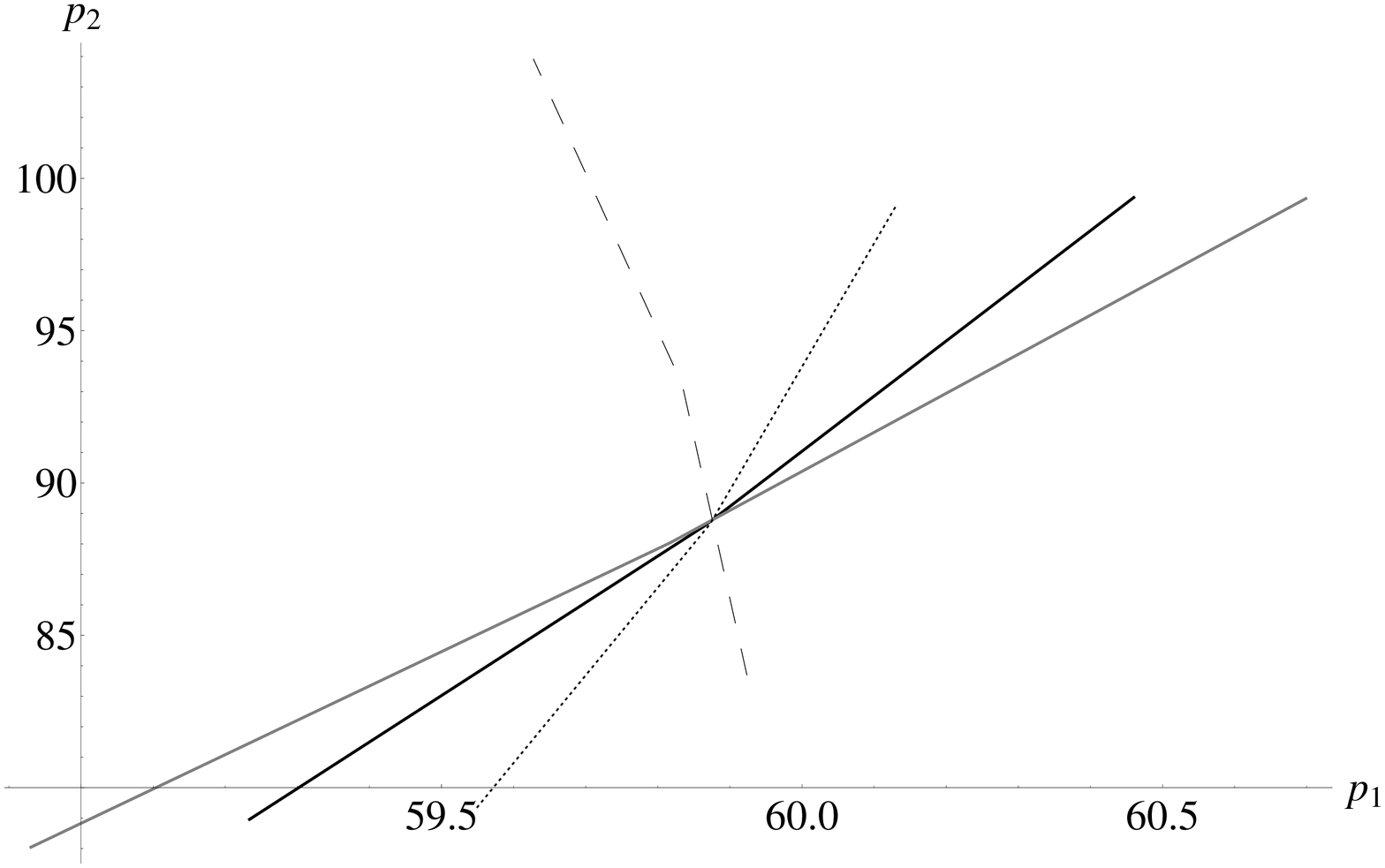}% Here is how to import EPS art
	\caption{\label{Md1pi2}Magnification of part of the graphic (Fig. \ref{d1pi2})}
\end{figure}

\begin{figure}[h]
	\includegraphics[width=\columnwidth]{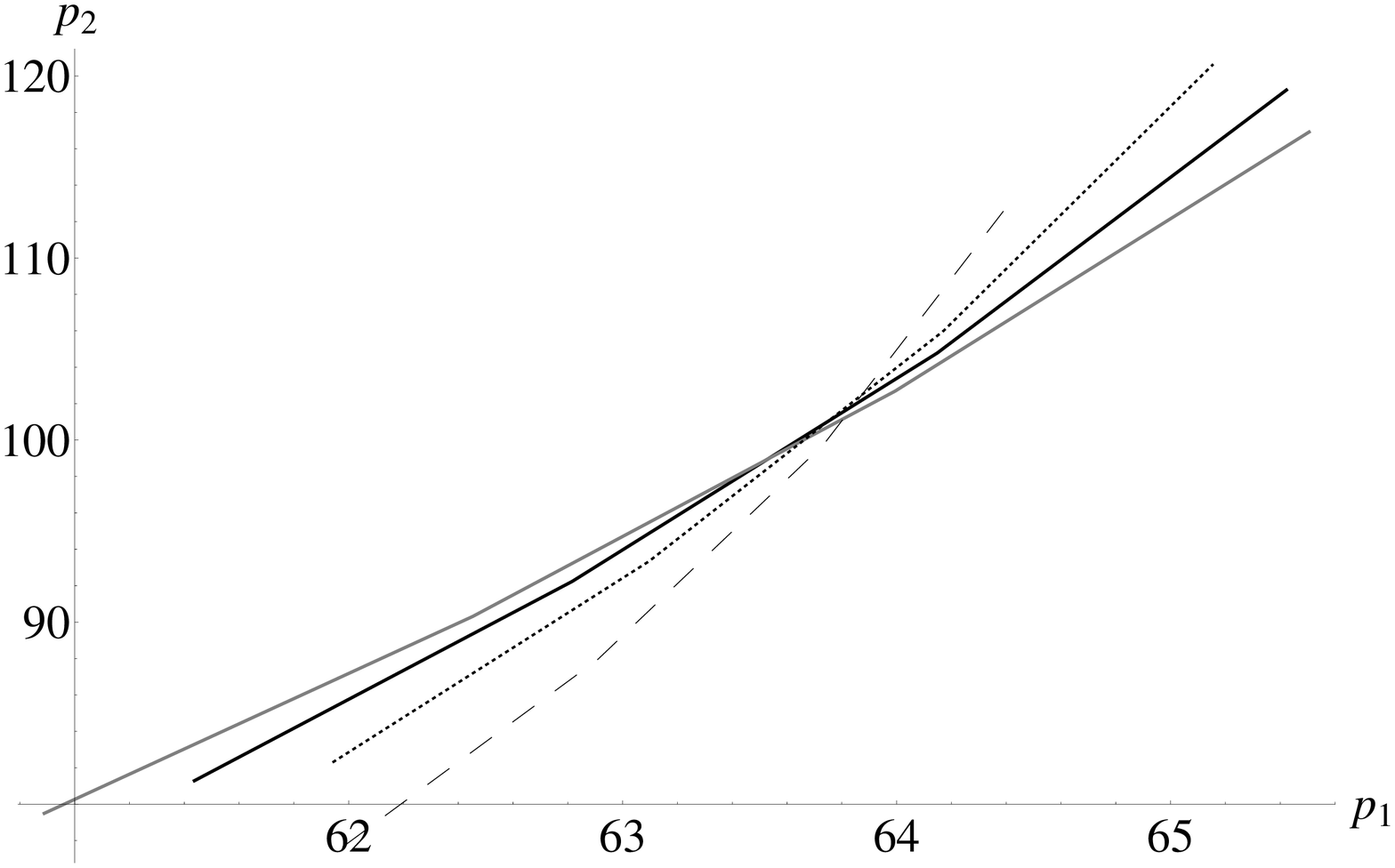}% Here is how to import EPS art
	\caption{\label{Md14pi2}Magnification of part of the graphic (Fig. \ref{d14pi2})}
\end{figure}

The main purpose of the present subsection is to obtain relation between integrals of motion of the star from one hand and the redshift $z$ and the derivative $\mathrm{d}z/\mathrm{d}\tau$ for all moments of proper time from another.
From (\ref{1deriv}) and (\ref{2deriv}) obtain:
\begin{equation}\label{FirstEq}
\EuScript{F}_1(r,\,z,\,E,\,L,\,i,\,\tilde{\theta})=\EuScript{F}_2(r,\,z,\,\frac{\mathrm{d}z}{\mathrm{d}\tau},\,E,\,L,\,i,\,\tilde{\theta})\,.
\end{equation} 
The equation (\ref{FirstEq}) gives possibilities to obtain connection between constant parameters of motion $L$, $E$ and $i$ in the case, where radius of radiation $r$ and the angle $\tilde\theta$ are known. Therefore, for the solution of the problem it is necessary to use more equations. For this purpose equations (\ref{1deriv}) and (\ref{BVPeq}) can be used. For this purpose express $\tilde{\theta}$ from (\ref{BVPeq}). Impact parameter $D$ in equation (\ref{BVPeq}) can be expressed using (\ref{1deriv}). Then obtain:
\begin{equation}\label{thetaz}
\tilde{\theta}=f(r,\,\EuScript{F}_1(r,\,z,\,E,\,L,\,i,\,\tilde{\theta}))\,.
\end{equation}
Here $f$ is the certain known function. This equation can be solved relative to $\tilde{\theta}$ using the iteration method. Then, obtain equation:
\begin{equation}\label{SecondEq}
r=r_s(\varphi(\tilde{\theta}(r,\,z,\,E,\,L,\,i),\,i)+\delta,\,E,\,L)\,.
\end{equation}
System of equations (\ref{FirstEq}), (\ref{SecondEq}) that is obtained, has 2 equations and 5 unknown variables. Because of this the solution of the system represents 3-surface in certain mathematical space (points of this space can be parametrized by variables $E$, $L$, $i$, $\delta$). This surface can be determined numerically. The results of calculations for our numerical model of the radiation of the star are presented  (see Fig. \ref{d1pi2}-\ref{Md14pi2}) for each point of redshift data. 
This is seen from Fig. \ref{d1pi2}-\ref{d14pi2} that for each point of data obtained 3-surfaces are not coincide. Therefore, points of intersection of the surfaces gives exact value of integrals of motion of the star. This point can be determined from obtained graphics with high accuracy.

Figures \ref{d1pi2}-\ref{Md14pi2} also illustrate that in the case where the angular parameters are chosen coincident with exact solution, sections have unique point of intersection (Fig. \ref{Md1pi2}), that is correspond to the solution of inverse problem. While for the not exact angular parameters, unique point of intersection does not exists (Fig. \ref{Md14pi2}). From the consideration of such graphics, it is possible to make a decision, that in the certain region of the surface exist only one point of intersection. Therefore, initial approximation to the exact solution of inverse problem can be found from such graphics with quite good accuracy. 

\subsection{Solution of inverse problem}
The farther improvement of the results can be obtain using the least squares method. For this purpose use the functions $\tau_s(\varphi,\,E,\,L)$ and $r_s(\varphi,\,E,\,L)$. From (\ref{FirstEq}) obtain the following equation:
\begin{eqnarray}
&&\EuScript{F}_1(r_s(\varphi+\delta,\,E,\,L),\,z,\,E,\,L,\,i,\,\tilde{\theta}(\varphi,\,i))=\nonumber\\
&&\EuScript{F}_2(r_s(\varphi+\delta,\,E,\,L),\,z,\,\frac{\mathrm{d}z}{\mathrm{d}\tau},\,E,\,L,\,i,\,\tilde{\theta}(\varphi,\,i))\label{LSMeq}\,.
\end{eqnarray}  
Left hand side of (\ref{LSMeq}) and right hand side of (\ref{LSMeq}) represents certain functions of $\varphi$. It is follows from (\ref{LSMeq}), that these functions must be equal for certain set of unknown parameters $E$, $L$, $i$, $\delta$. Therefore, mentioned parameters can be found using the least squares method. For this purpose in the presented example 10 points for different values of $\varphi$ in the range $[1 rad,\,1.4 rad]$ are chosen. Initial approximation obtain using the results of previous subsection. Obtained results presented in Table \ref{LSM_results}.
\begin{table}[h!]
	\caption{\normalsize\normalcolor Results \label{LSM_results}} 
	\begin{tabular}{l c c c}
		\hline
		Parameter & Initial  &  Reconstructed & Exact \\
		         & value     &  value       & value \\
		         &           &    \\
		\hline
		Pericenter distance, $p_1/M$ & 62,0 & 60,1 & 60,0 \\
		Apocenter distance, $p_2$ & 95,0 & 89,1 & 90,0\\
		Orbital inclination, $i$ & 1,5 rad & 1,48 rad & 1,4 rad\\
		Initial phase, $\delta$ & 0,9 rad & 1,0 rad & 1,0 rad \\
		\hline
	\end{tabular}
\end{table}    

\section{Conclusion}
The presented approach gives possibilities to solve the inverse problem: reconstruction of motion of a star moving in external gravitational field of supermassive black hole from redshift of the spectrum of received radiation. The presented approach uses properties of congruences of isotropic geodesics in order to formulate problem in terms of solutions of certain usual non linear equations that have form of analytical functions of integrals of motion of the star. The obtain equations are exact equations in General Theory of Relativity (we neglect only by optical scalar $\sigma$). Due to this the presented approach can be used for all possible sources (if they are can be considered as test particles in external gravitational field of black hole) moving on arbitrary distance to the black hole.

Apart from this the approach can be directly used in the case of the data of timing of pulsar, moving in external gravitational field. It is known, that large amount of pulsars anticipated to be detected in the Galactic Center in near future (see, e. g. \cite{Zhang2017}). Pulsars can move in the more close vicinity to supermassive black hole and due to this they are even more interesting for the purpose of testing theories of gravity. It is follows from general formula for the redshift (\ref{zform}), that times of arrival of pulses of pulsar can be expressed through redshift from the following equality:
\begin{equation*}
t_{TOA}^{(N)}=t_{TOA}^{(N-1)}+T_p(z+1)=t_{TOA}^{(N-1)}+T_p\frac{(k^iu_i)_s}{(k^iu_i)_o}\,.
\end{equation*}
Here $T_p$ --- is the pulsar period in the reference frame of the pulsar,  $z$ is the redshift, $t_{TOA}^{(j)}$ --- is the time of arrival of the $j$ th pulse. In the problem of reconstruction of motion of pulsar in the neighbourhood of supermassive black hole exist one more unknown parameter --- $T_p$. 

Another interesting application of the results of this paper is possibility of reconstruction of motion of a binary star in the vicinity of black hole. Knowledge of motion of such objects is very important problem in astrophysics and stellar mechanics (see, e. g. \cite{Kozai-Lidov,KLMechanism,KLMechanism2,BinarySBH,GW_BinarySBH}). The approach for the solution of the problem for the case of determination of relative motion of components only, is presented in the previous paper of the authors \cite{IJMPA2020}.

The approach can be used directly to the redshift data for the stars, moving in close vicinity to the supermassive black hole in the Galactic Center (for example, for S62 star \cite{S62_Star}) for the purpose of testing General Theory of Relativity.  For this it is possible to obtain integral of motion of a star from the presented algorithm and to calculate redshift as function of time of observation for future moments of time. Comparison of obtained curve with observational data will give the result of testing. With certain generalizations the approach can be used for the reconstruction of motion of a star in the vicinity of rotating black hole. We leave this problem for a future work.      

\appendix

\section{Solution of equations  (\ref{1deriv}) and (\ref{2deriv}) relative to $D/r$}
\begin{figure}[h]
	\includegraphics[width=\columnwidth]{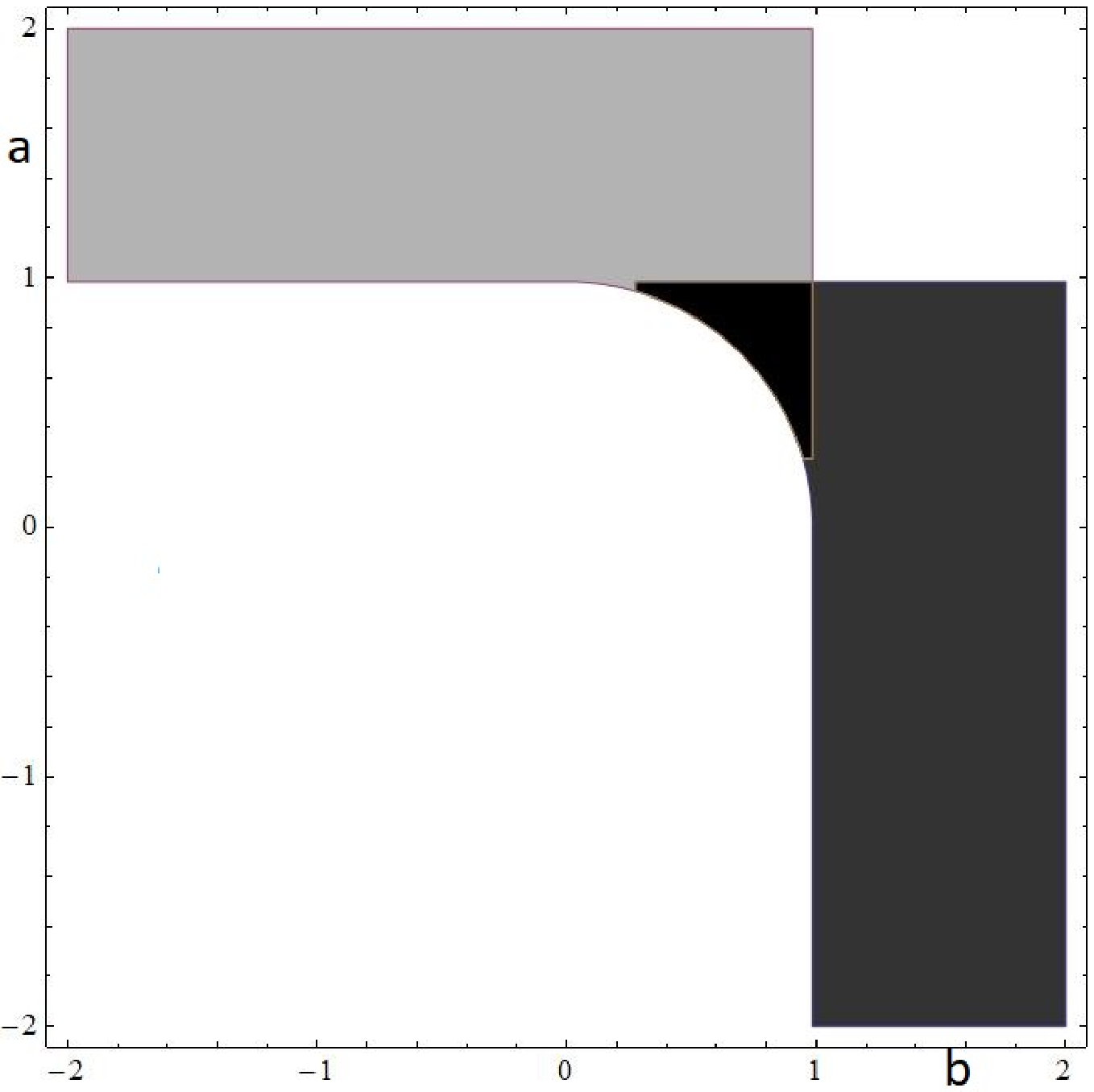}% Here is how to import EPS art
	\caption{\label{regions}Value $c=1,032$. Light grey region correspond to the region for sign $"+"$, dark grey region correspond to sign $"-"$ (see Table \ref{Solution}). Black region correspond to the intersection of both mentioned regions.}
\end{figure}

%\prepdef\appendix{}
In order to determine the unique signs in the equations (\ref{1deriv}) and (\ref{2deriv}) it is necessary to consider general case of such equation relative to $x$:
\begin{equation}
1=bx+a\sqrt{c-x^2}\,.\label{GenEq}
\end{equation}
From (\ref{GenEq}) one can obtain the following quadratic equation:
\begin{equation}
x^2(a^2+b^2)-2bx+1-a^2 c=0\,.\label{QuadEq}
\end{equation}
Two solutions of (\ref{QuadEq}) have the following form:
\begin{equation}
x=\frac{b^2}{a^2+b^2}\pm a\frac{\sqrt{c(a^2+b^2)-1}}{a^2+b^2}\,.\label{QSol}
\end{equation}
Which expression ($"+"$ or $"-"$) from (\ref{QSol}) is the solution of (\ref{GenEq}) can be found by direct substitution of (\ref{QSol}) to (\ref{GenEq}). Also take into account that solution $x=D/r$ mast be real and non-negative. Solution has sign $"+"$ or $"-"$ if all corresponding inequalities are hold (see Table \ref{Solution}). 
\begin{table}[h!]
	\caption{\normalsize\normalcolor Solution of equation (\ref{GenEq}) \label{Solution}} 
	\begin{tabular}{c c}
		\hline
		& \\
		$"+"$ & $"-"$ \\
		\hline
		& \\
		$c(a^2+b^2)-1\geqslant 0$ & $\qquad c(a^2+b^2)-1\geqslant 0$ \\
		$a-b\sqrt{c(a^2+b^2)-1}\geqslant 0$ & $\qquad a+b\sqrt{c(a^2+b^2)-1}\geqslant 0$ \\
		$b+a\sqrt{c(a^2+b^2)-1}\geqslant 0$ & $\qquad b-a\sqrt{c(a^2+b^2)-1}\geqslant 0$ \\
		\hline
	\end{tabular}
\end{table} 
The regions that are correspond to the inequalities in Table \ref{Solution} are depicted on the Figure \ref{regions}. It is follows from (\ref{1abc}) and (\ref{2abc}) that parameters $c$ are equal in both cases and they are slightly more than $1$ for the case of S-stars. Due to this we fix the typical value of the parameter $c=1,03$ on the graphic \ref{regions}.
Equation (\ref{zSchwarzschild}) relative to $D/r$ has the form of (\ref{GenEq}) with coefficients:
\begin{eqnarray}
&& a=-\frac{e_r e_s\sqrt{\frac{E^2}{1-2M/r}-\left(1+\frac{L^2}{r^2}\right)}}{1+z-\frac{E}{1-2M/r}}\,;\notag\\
&& b=\frac{L\beta}{r\left(1+z-\frac{E}{1-2M/r}\right)}\,;\notag\\
&& c=\frac{1}{1-2M/r}\,.\label{1abc}
\end{eqnarray}
Equation (\ref{zderivativeRes}) relative to $D/r$ also has the form of (\ref{GenEq}) with coefficients:
\begin{eqnarray}
&& a=e_r\sqrt{1-\frac{2M}{r}}\frac{\left(\frac{(1+z)^2}{r^2}-\frac{|A|^2}{r}\right)}{\mathrm{d}z/\mathrm{d}\tau}\,;\notag\\
&& b=\frac{\left(\ctan\tilde\theta |A|^2/r-2|A|\sin P_A (z+1)/r^2\right)}{\mathrm{d}z/\mathrm{d}\tau}\,;\notag\\
&& c=\frac{1}{1-2M/r}\,.\label{2abc}
\end{eqnarray}	
If parameters $a\, b\, c$ lie in the light grey or in the black grey region on Figure \ref{regions}, then equations (\ref{zSchwarzschild}) and (\ref{zderivativeRes}) have unique solution (\ref{1deriv}) and (\ref{2deriv}) respectively with corresponding sign. There fore we are firstly interesting to find solution in these regions. If all numerical solutions lie in black region of Figure \ref{regions}, it is necessary to consider two possible cases with signs $"+"$ and $"-"$ in (\ref{1deriv}) - (\ref{2deriv}). Then solution with minimal $\chi^2$ must be chosen.   

%\bibliographystyle{apsrev4-2}
%\bibliography{Schwarzschild_Inverse}% Produces the bibliography via BibTeX.
\end{document}